\documentclass[aip,graphicx,reprint]{revtex4-1}
\usepackage{graphicx}% Include figure files
\usepackage{dcolumn}% Align table columns on decimal point
\usepackage{bm}% bold math

\draft % marks overfull lines with a black rule on the right

\begin{document}




%\preprint{}

%\title{A force field of main-chain dihedral angles for each amino acid in protein systems} %Title of paper
\title{A Conformational Search Method for Protein Systems Using Genetic Crossover and Metropolis Criterion} %Title of paper

\author{Yoshitake Sakae}
\affiliation{Department of Physics, Graduate School of Science, Nagoya University, 
Nagoya, Aichi 464-8602, Japan}
\author{Tomoyuki Hiroyasu}
\affiliation{Department of Biomedical Information, Doshisha University, Kyotanabe, Kyoto 610-0394, Japan}
\author{Mitsunori Miki}
\affiliation{Department of Intelligent Information Engineering and Sciences, Doshisha University, Kyotanabe, Kyoto 610-0394, Japan}
\author{Katsuya Ishii}
\affiliation{Information Technology Center, Nagoya University, Nagoya, 
Aichi 464-8601, Japan}
\author{Yuko Okamoto}
\affiliation{Department of Physics, Graduate School of Science, Nagoya University, 
Nagoya, Aichi 464-8602, Japan}
\affiliation{Information Technology Center, Nagoya University, Nagoya, 
Aichi 464-8601, Japan}
\affiliation{Structural Biology Research Center, Graduate School of Science, Nagoya University, Nagoya, Aichi 464-8602, Japan}
\affiliation{Center for Computational Science, Graduate School of Engineering, Nagoya University, Nagoya, Aichi 464-8603, Japan}








% \author{}



%\thanks{}


% \affiliation{}





%\date{\today}

\begin{abstract}
Many proteins carry out their biological functions by forming the characteristic tertiary structures.
Therefore, the search of the stable states of proteins by molecular simulations is important to understand their functions and stabilities.
However, getting the stable state by conformational search is difficult,
because the energy landscape of the system is characterized by many local minima separated by high energy barriers.
In order to overcome this difficulty, various sampling and optimization methods for conformations of proteins have been proposed.
In this study, we propose a new conformational search method for proteins by using genetic 
crossover and Metropolis criterion.
We applied this method to an $\alpha$-helical protein.
The conformations obtained from the simulations are in good agreement with the experimental results.
\end{abstract}


\maketitle %\maketitle must follow title, authors, abstract and \pacs



\section{Introduction}
To understand functions and stabilities of biomolecules such as DNA and proteins, 
effective conformational search and accurate estimation of canonical distributions are important.
However, because biomolecules have a lot of local minimum-energy states separated by high energy barriers, 
conventional molecular dynamics (MD) and Monte Carlo (MC) simulations tend to get trapped in states of 
local minima.
In order to overcome this difficulty, 
various sampling and optimization methods for conformations of biomolecules have been proposed 
such as generalized-ensemble algorithms (for reviews, 
see, e.g., \cite{GEA1,GEA2}). 
%which include the multicanonical (MUCA) \cite{MUCA1,MUCA2}, the 
%simulated-tempering (ST) \cite{ST1,ST2} 
%and the replica-exchange (REM) \cite{REM1,REM2}.
Moreover, we have proposed a conformational search method called the parallel simulated 
annealing using genetic crossover (PSA/GAc) \cite{PSAGAc,PSAGAc2,PSAGAc3,PSAMDGAc}, 
which is a hybrid algorithm combining both the simulated annealing (SA) \cite{SA} 
and the genetic algorithm (GA) \cite{GA1,GA2}.
In this method, a simulated annealing simulation is combined 
with genetic crossover, 
which is one of the operations of genetic algorithm.
% The conformational search using simulated annealing 
% is based on local conformational updates. 
% On the other hand, the genetic algorithm is based on 
% global conformational updates.
% Our method incorporates these two attractive features of 
% the simulated annealing and the genetic crossover.
% We compared our results
% with those of the normal simulated annealing molecular dynamics
% simulations by using an $\alpha$-helical miniprotein. 
% As the results, the conformations which have lower energy than those obtained from the
% conventional simulated annealing were obtained.
% Moreover, the results showed a possibility that a combination method with the operation 
% of the genetic crossover would be useful for the protein systems.
% Moreover, the various conformational search or optimization approaches for 
% biomolecules using the genetic algorithm 
% have also been 
% performed\cite{GA-protein1,GA-protein2,GA-protein3,GA-protein4,GA-protein5}.

In this paper, we propose a new conformational search method 
%by using the genetic crossover and Metropolis criterion.
that combines the genetic crossover and Metropolis criterion \cite{Metro}.
The operation of the genetic crossover is combined with the conventional 
MC or MD simulations.
In order to examine the effectiveness of our method,
we applied our method to a small protein, which is the 10-55 helical fragment B of protein A 
from {\it Staphylococcus aureus} 
(in this paper, just referred to as protein A) \cite{proteinA}.
The results of this simulation are found to be in good agreement with the experimental results.

\section{Methods}
\label{method}
%We show the schematic process of our genetic crossover sampling method in Fig.~\ref{fig_schematic}. 
In the present simulation approach, 
we first prepare $M$ initial conformations of the system in study, 
where $M$ is the total number of ``individuals'' in genetic algorithm
and is usually taken to be an even integer.
We then alternately perform the following two steps:
\begin{enumerate}
\renewcommand{\labelenumi}{\arabic{enumi}.}
\item 
For the $M$ individuals, regular canonical MC or MD simulations 
at a fixed temperature $T$ are carried out simulataneously and independently 
for a certain MC or MD steps.
\item 
$M/2$ pairs of conformations are selected from ``parental'' 
group randomly, and the crossover and selection operations are performed.
Here, the parental group means the latest conformations obtained in Step 1.
\end{enumerate}
While Step 1 is usually based on local updates of conformations,
global updates of conformations are introduced in Step 2 by
genetic crossover.  The latter greatly enhances the conformational
sampling.
 
In the following, we give the details of Step 2 above.
We can employ various kinds of genetic crossover operations such as 
one-point crossover \cite{PSAGAc,PSAGAc2,PSAGAc3}, 
two-point crossover \cite{PSAMDGAc},
etc. as in our previous simulation methods.
Here, we just present a case of the two-point genetic crossover.
The crossover operation in this method exchanges a part of 
corresponding dihedral angles 
between two conformations of the protein.
In the two-point crossover operation on a parental pair, the following procedure is carried out:
% In the two-point crossover operation, the following procedure is 
%carried out (see Fig.~\ref{fig_crossover}) :\\
%
\begin{figure}
\begin{center}
\resizebox*{8cm}{!}{\includegraphics{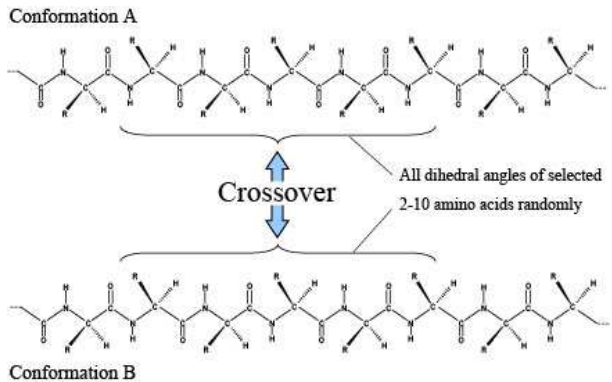}}%
\caption{Schematic process of the two-point crossover operation.
In this process, all dihedral angles (in backbone and side chains) within the randomly selected $n$ consecutive amino acids are exchanged 
between a pair of conformations.}
\label{fig_crossover}
\end{center}
\end{figure}

\begin{enumerate}
\item Consecutive amino acids of length $n$ residues 
in the amino-acid sequence of the conformation are selected 
randomly for each pair of selected conformations.
\item Dihedral angles (in only backbone or all dihedral angles) in 
the selected $n$ amino acids are exchanged between the selected 
pair of conformations. 
\end{enumerate}
Note that the length $n$ of consecutive amino-acid residues can, 
in general, be different for each pair of selected conformations.
  
We need to deal with the produced ``child'' conformations with care.
Because the produced conformations often have unnatural structures 
by the crossover operation, they 
have high potential energy and are unstable.
Therefore, the relaxation process is introduced before the selection operation.
% We propose two kinds of therelaxation process.
% One is to adding the energy minimization after exchanging dihedral angles of the conformations.
% For these two exchanged conformations, we perform the ``light'' energy minimizations 
% by a standard method such as Newton-Raphson method and 
% conjugate gradient method 
% in order to avoid any obvious unnatural parts of conformations.
% Additionally, the equilibrium simulations are performed.
% Another is to adding the equilibrium simulations with the restraints.
In this paper, we use equlibration simulations with restraints 
as a relaxation process. 
For instance, short simulations at the same temperature $T$
with restraints 
on the backbone dihedral angles of only the $n$ 
amino acids are performed so that the corresponding backbone
structures of the $n$ amino acids 
will approach the exchanged backbone conformation.
The initial conformations for these equilibration simulations
are the ones before the exchanges.
Namely, by these equilibration simulations,
the corresponding backbone conformations of the $n$ amino acids
gradually transform from the ones
before the exchanges to the ones after the exchanges. 
% We expect that this procedure increases the acceptance ratio of the crossover operation
% in comparison with the minimization procedure, especially, when the case of larger protein systems (data not shown).
%This procedure generates trial conformations by reaching the trial ``child'' conformation from
%the original ``parental'' conformation gradually.
We then perform short equilibration simulations without the restraints.
% In this method, as a simple way to monitor system equilibration, we record the values of the potential energy and temperature.
We select the last conformations in the equilibratoin simulations
as ``child'' conformations.
Note that from a pair of parental conformations we get two child conformations.
In the present method, we consider selection between parent and child
for the parent-child pair with the same conformations in the remaining
amino acids (other than $n$ consecutive ones).
  
In the final stage in Step 2, the selection operation is performed.
We select a superior ``chromosome'' (conformation)
from the parent-child pair.
For this selection operation, we can also employ various criteria.
In this study, we employ Metropolis criterion,
which selects the new child conformation from the parent with 
the following probability:
\begin{equation}
w({\rm p} \rightarrow  {\rm c}) = {\rm min} \left( 1, {\rm exp}\{-\beta [ E_{\rm c} - E_{\rm p} ] \}  \right).
\end{equation}
Here, $E_{\rm p}$ and $E_{\rm c}$ stand for the potential energy of the parental conformation and 
the child conformation, respectively, of the parent-child pair.
$\beta$ is the inverse temperature, which is defined by $\beta = 1/{k_{\rm B} T}$ 
($k_{\rm B}$ is the Boltzmann constant).

% We expected to obtain the effective sampling data around the energy space by using this operation.

We remark that if we use different temperatures,
we can also introduce the replica-exchange method \cite{REM1,REMD} 
together with the above method
in order to further enhance conformational sampling \cite{SHMIO}.

\section{Results and Discussion}
We applied the present method to protein A.
Although the whole protein A has 60 amino acids, 
we used the truncated 46 amino-acid sequence from Gln10 to Ala55. 
For this simulation, we used the AMBER12 program package and incorporated the 
two-point genetic crossover procedure.
The unit time step was set to 2.0 fs and the bonds involving hydrogen atoms
were constrained by the SHAKE algorithm \cite{shake}.
Each simulation for sampling was carried out for 90.0 nsec (which consisted 
of 45,000,000 MD steps) with 32 individuals 
($M = 32$) and performed the crossover operations 90 times
during the simulation.
The temperature during the simulations was kept at 300 K by using Langevin dynamics.
The nonbonded cutoff of 20 \AA~ was used.
As for solvent effects, we used the GB/SA model \cite{gbsa_igb5} included
in the AMBER12 program package ($igb = 5$).
In the crossover operations, 
we set the length $n$ of consecutive amino-acid residues to be an even integer 
ranging from 10 to 20.  This number was chosen randomly for each pair of
parental conformations.
As for the equilibration simulations just after each genetic crossover
operation,
the first simulations with the harmonic restraints on the backbone
diheral angles of $n$ amino-acid residues (the force constants
were 600 kcal/mol$\cdot$\AA$^2$)
lasted for 20 psec, and the following simulations without restraints
also lasted for 20 psec.
%Two kinds of the equilibrium simulations were performed for the trial conformations.
% The first equilibrium is with the restraints for the backbone-dihedral angles of 
% the exchanged amino acids instead of the minimizations at the above simulations of Trp-cage
% (see the parts of the relaxation process in Methods section). 

\begin{figure}
\begin{center}
\resizebox*{8cm}{!}{\includegraphics{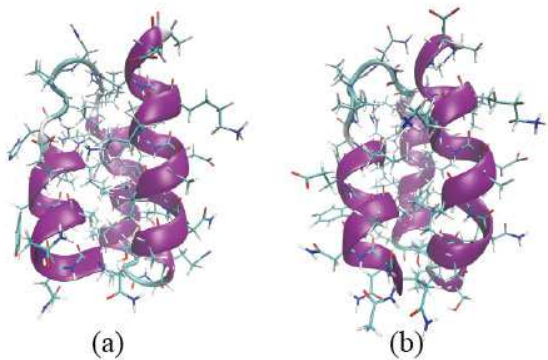}}%
\caption{Structures of protein A. 
(a) PDB structure (PDB ID: 1BDD). 
(b) A conformation obtained from the present simulation,
which has the lowest RMSD value from the PDB structure (RMSD = 1.7 \AA ).
}
\label{fig_pa_strs}
\end{center}
\end{figure}

\begin{figure}
\begin{center}
\resizebox*{8cm}{!}{\includegraphics{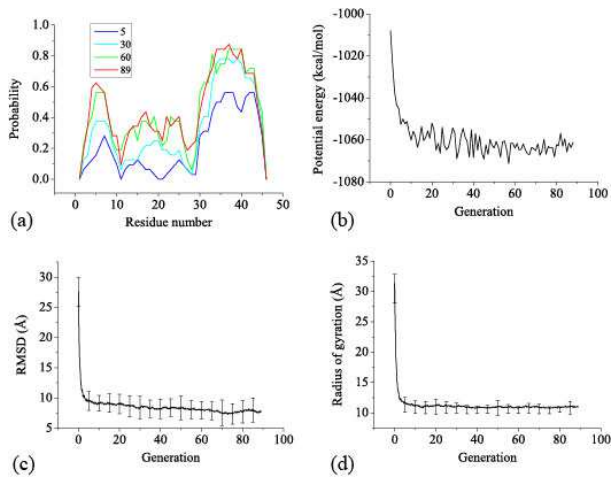}}%
\caption{(a) The fraction of helix conformations as a function
of residue number at generation numbers 5 (blue), 30 (light blue),
60 (green), and 89 (red), (b) the average of the potential energy, 
(c) the average of the RMSD,
and (d) the average of the radius of gyration, obtained from the 
present GC simulation results.
The helix conformations were examined by using the DSSP program \cite{DSSP}. 
The error bars in (c) and (d) are 
standard deviations.}
\label{fig_generation_values}
\end{center}
\end{figure}

We obtained a similar conformation to the experimental native structure,
and its root-mean-square distance (RMSD) (for only the backbone atoms)
from the native structure was
1.7 \AA~ (see Fig.~\ref{fig_pa_strs}).
One of the characteristic analyses of the genetic algorithm is the analysis of data 
as functions of the generation number.
In this paper, the generation number stands for the number of crossover operations.
In Fig.~\ref{fig_generation_values}, we show four kinds of values, the fraction of 
helix conformations, 
the average of the potential energy, the average of the RMSD, and the average of the radius of gyration 
by changing the generation.
In Fig.~\ref{fig_generation_values}(a), we see that the fraction of helix conformations increases 
as the generation proceeds as a whole.
The PDB structure of protein A has three helix regions in the amino-acid sequence 
of 2--9, 17--29, and 34--46. 
%by using DSSP program \cite{DSSP}.
These helix regions roughly correspond to the high probability regions 
obtained from the present simulation, and the C-terminal 
region (Helix III) has the 
highest helix fraction among the three regions
 in Fig.~\ref{fig_generation_values}(a).
The stabilities for the three helix structures in protein A have been examined and debated by many researchers 
\cite{proteinA_Zhou_Karplus,proteinA_Alonso_Daggett,proteinA_Garcia_Onuchic,proteinA_Sato_etal,proteinA_Shao_Gao}.
>From our simulation results, we consider that Helix III is the most stable, 
and the central helix is unstable 
in comparison with the other two helix structures.
In Fig.~\ref{fig_generation_values}(b)--(d), we see that
the potential energy, RMSD, and radius of gyration quickly decrease 
at the early numbers of generations.
After that, although these values are roughly convergent, there are some fluctuations.
For example, RMSD value has the standard deviation of 1.7 \AA~ around a mean
value of about 8.0 \AA.

\section{Conclusions}
\label{conclusions}
In this work, we proposed a new conformational search method for protein systems.
This method combines conventional canonical MC or MD simulations, 
which are based on local updates of conformations, and
genetic crossover, which is 
based on global updates of conformations.
The latter greatly enhances the conformational sampling.
In order to examine the efficiency of this method, we applied it
to protein A.
We found conformations very close to the native structure.
%However, the conformations did not have the global minimum during the simulation.
%We consider that one of the reasons is the instability of Helix II in the three helix structures.

The present method is particularly suitable for highly
parallel computers.
In the future, we are going to apply this method to larger protein systems.

\section*{Acknowledgements}
The computations were performed on the computers at the Research Center for Computational Science, 
Institute for Molecular Science, Information Technology Center, Nagoya University, and Center for 
Computational Sciences, University of Tsukuba.
This work was supported, in part, by 
the Grants-in-Aid,
for Scientific Research on Innovative Areas (``Dynamical Ordering \& Integrated Functions''),
for the Computional Materials Science Initiative,
and for High Performance Computing Infrastructure, 
from the Ministry of Education, Culture, Sports, Science and Technology (MEXT), Japan.

%\label{}

% \subsection{}

% \subsubsection{}

% If in two-column mode, this environment will change to single-column format so that long equations can be displayed. 


%\begin{widetext}

%$$\mbox{put long equation here}$$

%\end{widetext}

% Figures should be put into the text as floats. 

% Use the graphics or graphicx packages (distributed with LaTeX2e).


%



% Put the label that you will use with \ref{} command in the braces of the \label{} command.

%

% \begin{figure}



% \end{figure}







%

% \begin{table}



% \end{tabular}

% \end{table}


~~\\
~~\\

\noindent
{\bf REFERENCES}

% \bibliographystyle{aipauth4-1}
% \bibliography{jcp2011}
%\bibliography{jpcb2012}

\begin{thebibliography}{10}
\makeatletter
\providecommand \@ifxundefined [1]{%
 \@ifx{#1\undefined}
}%
\providecommand \@ifnum [1]{%
 \ifnum #1\expandafter \@firstoftwo
 \else \expandafter \@secondoftwo
 \fi
}%
\providecommand \@ifx [1]{%
 \ifx #1\expandafter \@firstoftwo
 \else \expandafter \@secondoftwo
 \fi
}%
\providecommand \natexlab [1]{#1}%
\providecommand \enquote  [1]{``#1''}%
\providecommand \bibnamefont  [1]{#1}%
\providecommand \bibfnamefont [1]{#1}%
\providecommand \citenamefont [1]{#1}%
\providecommand \href@noop [0]{\@secondoftwo}%
\providecommand \href [0]{\begingroup \@sanitize@url \@href}%
\providecommand \@href[1]{\@@startlink{#1}\@@href}%
\providecommand \@@href[1]{\endgroup#1\@@endlink}%
\providecommand \@sanitize@url [0]{\catcode `\\12\catcode `\$12\catcode
  `\&12\catcode `\#12\catcode `\^12\catcode `\_12\catcode `\%12\relax}%
\providecommand \@@startlink[1]{}%
\providecommand \@@endlink[0]{}%
\providecommand \url  [0]{\begingroup\@sanitize@url \@url }%
\providecommand \@url [1]{\endgroup\@href {#1}{\urlprefix }}%
\providecommand \urlprefix  [0]{URL }%
\providecommand \Eprint [0]{\href }%
\providecommand \doibase [0]{http://dx.doi.org/}%
\providecommand \selectlanguage [0]{\@gobble}%
\providecommand \bibinfo  [0]{\@secondoftwo}%
\providecommand \bibfield  [0]{\@secondoftwo}%
\providecommand \translation [1]{[#1]}%
\providecommand \BibitemOpen [0]{}%
\providecommand \bibitemStop [0]{}%
\providecommand \bibitemNoStop [0]{.\EOS\space}%
\providecommand \EOS [0]{\spacefactor3000\relax}%
\providecommand \BibitemShut  [1]{\csname bibitem#1\endcsname}%
\let\auto@bib@innerbib\@empty
%</preamble>
\bibitem{GEA1}
Hansmann U and Okamoto Y, {\em Curr. Opin. Struct. Biol.\/} {\bf 9}, 
  177--183 (1999).

\bibitem{GEA2}
Mitsutake A, Sugita Y and Okamoto Y, {\em Biopolymers\/} {\bf 60}, 96--123 (2001).

\bibitem{PSAGAc}
Hiroyasu T, Miki M and Ogura M, {\em Proceedings of the 44th Institute of
  Systems\/} 2000, pp. 113--114.

\bibitem{PSAGAc2}
Hiroyasu T, Miki M, Ogura M and Okamoto Y, {\em J. Inf. Proc. Soc. Jpn.\/} {\bf 43},
  70--79 (2002).

\bibitem{PSAGAc3}
Hiroyasu T, Miki M, Ogura M, Aoi K, Yoshida T and Okamoto Y, {\em
  Proceedings of the 7th World Multiconference on Systemics, Cybernetics and
  Informatics (SCI 2003)\/}, pp. 117--122 (2003).

\bibitem{PSAMDGAc}
Sakae Y, Hiroyasu T, Miki M and Okamoto Y, {\em J. Comput. Chem.\/} {\bf
  32}, 1353--1360 (2011).

\bibitem{SA}
Kirkpatrick S, Gelatt~Jr C D and Vecchi M P, {\em Science\/} {\bf 220}, 
  671--680 (1983).

\bibitem{GA1}
Holland J H, {\em Adaptation in Natural and Artificial Systems\/} (Ann
  Arbor: The University of Michigan Press, 1975).

\bibitem{GA2}
Goldberg D E, {\em Genetic Algorithms in Search, Optimization, and Machine
  Learning\/} (Reading: Addison-Wesley, 1989).
   
\bibitem{Metro}
Metropolis N, Rosenbluth A W, Rosenbluth M N, Teller A H, and
Teller E, {\em J. Chem. Phys.\/} {\bf 21}, 1087--1092 (1953).
   
\bibitem{proteinA}
Gouda H, Torigoe H, Saito A, Sato M, Arata Y and Shimada I, {\em
  Biochemistry\/} {\bf 31}, 9665--9672 (1992).

\bibitem{REM1}
Hukushima K and Nemoto K, {\em J. Phys. Soc. Jpn.\/} {\bf 65}, 1604--1608 (1996).

\bibitem{REMD}
Sugita Y and Okamoto Y, {\em Chem. Phys. Lett.\/} {\bf 314}, 141--151 (1999).

\bibitem{SHMIO}
Sakae Y, Hiroyasu T, Miki M, Ishii K and Okamoto Y, {\em Mol. Simul.\/} 
{\bf 41} (2015), in press.

\bibitem{shake}
Ryckaert J P, Ciccotti G and Berendsen H J C, {\em J. Comput. Phys.\/} {\bf
  23}, 327--341 (1977).

\bibitem{gbsa_igb5}
Onufriev A, Bashford D and Case D A, {\em Proteins\/} {\bf 55}, 383--394 (2004).

\bibitem{proteinA_Zhou_Karplus}
Zhou Y and Karplus M, {\em Nature\/} {\bf 401}, 400--403 (1999).

\bibitem{proteinA_Alonso_Daggett}
Alonso D O V and Daggett V, {\em Proc. Natl. Acad. Sci. U.S.A.\/} {\bf 97},
  133--138 (2000).

\bibitem{proteinA_Garcia_Onuchic}
Garcia A E and Onuchic J N, {\em Proc. Natl. Acad. Sci. U.S.A.\/} {\bf 100},
  13898--13903 (2003).

\bibitem{proteinA_Sato_etal}
Sato S, Religa T L, Daggett V and Fersht A R, {\em Proc. Natl. Acad. Sci.
  U.S.A.\/} {\bf 101}, 6952--6956 (2004).

\bibitem{proteinA_Shao_Gao}
Shao Q and Gao Y Q, {\em J. Chem. Phys.\/} {\bf 135}, 135102 (2011).

\bibitem{DSSP}
Kabsch W and Sander C, {\em Biopolymers\/} {\bf 22}, 2577--2637 (1983).
\end{thebibliography}


%\vspace{2cm}

~~\\
~~\\
~~\\
~~\\
~~\\
~~\\
~~\\
~~\\
~~\\
~~\\

%\newpage

%\begin{figure}
%\begin{center}
%\resizebox*{8cm}{!}{\includegraphics{crossover.eps}}%
%\caption{Schematic process of the two-point crossover operation.
%In this process, all dihedral angles (in backbone and side chains) within the randomly selected $n$ %consecutive amino acids are exchanged 
%between a pair of conformations.}
%\label{fig_crossover}
%\end{center}
%\end{figure}

%\begin{figure}
%\begin{center}
%\resizebox*{8cm}{!}{\includegraphics{min-rmsd-strs.eps}}%
%\caption{The lowest-RMSD structures obtained from the 
%conventional MD (Canonical), PMD/GAc (GC), conventional
%REMD (REM), and REMD/GAc (GC/REM).
%(a) is the lowest-RMSD structures with respect to the experimental 
%result (Native(PDB), PDB ID:1L2Y model1),
%and (b) is the lowest-RMSD structures with respect to the lowest-energy 
%conformation obtained from the conventional iso-thermal
%canonical simulations at 282 K (Native-Like(282K)).}
%\label{fig_min_rmsd_strs}
%\end{center}
%\end{figure}

%\begin{figure}
%\begin{center}
%\resizebox*{8cm}{!}{\includegraphics{P-rmsd.eps}}%
%\caption{Probability distributions of RMSD values of all conformations 
%\textcolor{red}{at temperature 282 K,} 
%obtained from the conventional MD (Canonical), 
%PMD/GAc (GC), conventional
%REMD (REM), and REMD/GAc (GC/REM).}
%\label{fig_P_rmsd}
%\end{center}
%\end{figure}

\end{document}